\documentclass[%
twocolumn,
amsmath,amssymb,
aps, physrev,
]{revtex4-2}

\usepackage{graphicx}% Include figure files
\usepackage{dcolumn}% Align table columns on decimal point
\usepackage{bm}% bold math
\usepackage{amsthm}
\usepackage{braket}
\newtheorem{theorem}{Theorem}
\newtheorem{corollary}{Corollary}[theorem]
\newtheorem{observation}{Observation}
\begin{document}
	
	\preprint{APS/123-QED}
	
	\title{\textbf{Nonlocal contributions to ergotropy: A thermodynamic perspective.} 
	}% 

	\author{B. Vigneshwar}
	\author{R. Sankaranarayanan}%
	\affiliation{Department of Physics, National Institute of Technology, Tiruchirappalli, 620015, TamilNadu, India\\}%Lines break automatically or can be forced with \\
	
	\date{\today}% It is always \today, today,
	
	\begin{abstract}
		Nonlocality is a defining feature of quantum mechanics and has long served as a key indicator of quantum resources since the formulation of Bell's inequalities. Identifying the contribution of nonlocality to extractable work remains a central problem in quantum thermodynamics. We address this by introducing a quantifier of nonlocal contributions to extractable work in bipartite systems. It is shown that closed form expressions can be calculated for our quantity in terms of the Schmidt coefficients. Further for strictly non-interacting Hamiltonian, the direct relationship between ergotropy and correlations is established. Our results reveal that nonlocal resources invariably enhance extractable work under non-interacting Hamiltonians, while in the presence of interactions, their contribution can either increase or diminish depending on the structure of the state and the Hamiltonian.
		
	\end{abstract}

	\maketitle
	
	\section{Introduction}
	Locality refers to the propagation of effects through direct physical processes bounded by the speed of light. Quantum mechanics, however, permits violations of locality via instantaneous correlations between entangled states \cite{bell1964physics}. Quantum nonlocality, which manifests as correlations that defy classical hidden variable theories, has been extensively studied through Bell inequalities \cite{einstein1935einsteinpodolskyrosen,werner2001bell,jones2007entanglement,horodecki2009quantum}.
	
	Quantum batteries aim to outperform their classical counterparts by exploiting quantum effects. Enhanced work capacity \cite{alicki2013entanglement,perarnau2015extractable,andolina2019extractable,yang2023battery,tirone2024quantum,vigneshwar2025noise}, fast charging \cite{hovhannisyan2013entanglement,binder2015quantacell,campaioli2017enhancing,ferraro2018high,gyhm2022quantum,shastri2025dephasing} and stabilization \cite{santos2019stable,farina2019charger,barra2019dissipative,gherardini2020stabilizing,quach2020using,song2024remote} have been studied using quantum batteries \cite{campaioli2024colloquium}. These theoretical developments have been supported by hardware simulations \cite{razzoli2025cyclic,gemme2022ibm} and experimental realizations \cite{quach2022superabsorption,hu2022optimal,maillette2023experimental}. Ergotropy \cite{allahverdyan2004maximal} was introduced to quantify the performance of quantum batteries. It is defined as the maximum unitarily extractable work from a system. 	
	Recent efforts have sought to connect ergotropy with quantum features like entanglement \cite{touil2021ergotropy,shi2022entanglement,francica2022quantum,gyhm2024beneficial,yang2024multiparticle,simon2025correlations} and coherence \cite{francica2020quantum,tirone2025quantum}. This has led to ergotropy-based quantifiers of correlations like ergodiscord \cite{alimuddin2025ergodiscord}. While much of these works focuses on state-based quantum correlations, the influence of the system Hamiltonian on ergotropy remains a key practical consideration. Frameworks such as the ergotropic gap \cite{mukherjee2016presence,alimuddin2019bound,puliyil2022thermodynamic} primarily assess quantum correlations in the state, often overlooking the Hamiltonian’s structural contributions to work extraction. Yet, since ergotropy depends on both the state and the Hamiltonian, understanding how interactions within the Hamiltonian modulate the role of correlations is crucial. Our objective is to develop a framework that captures the joint contribution of the Hamiltonian and the nonlocal resources in the battery state to work extraction.
	
	In this Letter, we study nonlocal resources from a quantum thermodynamic perspective using ergotropy, with the goal of elucidating quantum advantages in energy extraction. We introduce a quantity—Ergotropy-based Measurement-Induced Nonlocality (EMIN)—that characterizes when and how nonlocal correlations within a quantum battery contribute beneficially to ergotropy. Inspired by geometric measures of Measurement-Induced Nonlocality \cite{luo2011measurement,xi2012measurement,hu2015measurement,muthuganesan2017fidelity}, EMIN captures the ergotropy difference between the original state and the post-measurement state following a locally invariant projective measurement. A defining feature of EMIN is its dependence on the system Hamiltonian, setting it apart from traditional MIN-type measures.
	
	We begin by introducing EMIN and establishing its fundamental properties. We show the limitations of ergotropic gap as an entanglement measure in distinguishing product states under interacting Hamiltonian and establish that EMIN’s robustness overcomes this limitation. We then analyze non-interacting Hamiltonians and rigorously demonstrate that, in such cases, nonlocal correlations always enhance ergotropy by establishing correspondence between EMIN and geometric measures. We then generalize to arbitrary interacting Hamiltonians, revealing how specific interaction structures can either promote or hinder the ergotropic utility of nonlocal correlations. In summary, we establish that presence of correlation in the battery state can affect the ergotropy both positively and negatively depending on the energy landscape of the system. This unified framework highlights the intricate interplay between quantum correlations and Hamiltonian structure, offering new insights into the optimal use of nonlocal resources for quantum thermodynamic tasks.

	\section{Ergotropy-Based Measurement-Induced Nonlocality}
	Ergotropy quantifies the maximum work extractable from a quantum state via a cyclic unitary process \cite{allahverdyan2004maximal}:
	\begin{equation}
		\xi(\rho, H) = E(\rho) - E_p(\rho) = \operatorname{Tr}(\rho H) - \min_{V} \operatorname{Tr}(V \rho V^\dagger H),
		\label{e1}
	\end{equation}
	where \( E(\rho) \) is the energy of the state \( \rho \), and \( E_p(\rho) \) is the energy of its passive counterpart, obtained by minimizing over all unitary operations.
	
	Let \( \rho = \sum_n r_n \left|r_n\right\rangle\!\left\langle r_n\right| \) and \( H = \sum_n e_n \left|e_n\right\rangle\!\left\langle e_n\right| \) denote the spectral decompositions of the state and the Hamiltonian, respectively. The passive state relative to \( H \) is then \( \rho^p = \sum_n r_n \left|e_n\right\rangle\!\left\langle e_n\right| \), and the passive energy becomes
	\begin{equation}
		E_p(\rho) = \sum_n r_n^\downarrow e_n^\uparrow,
	\end{equation}
	where \( \{r_n^\downarrow\} \) and \( \{e_n^\uparrow\} \) denote the eigenvalues of \( \rho \) and \( H \), sorted in decreasing and increasing order, respectively.

	We define \textit{Ergotropy-based Measurement-Induced Nonlocality} (EMIN) as
	\begin{equation}
		N_\xi(\rho, H) = \max_{\Pi^a} \left[ \xi(\rho, H) - \xi\left(\Pi^a(\rho), H\right) \right],
		\label{e2}
	\end{equation}
	where \( \Pi^a \) denotes a set of local projective measurements \( \{\Pi_k^a = |k\rangle\!\langle k| \otimes \mathbb{I}_b\} \) that leave the marginal state \( \rho_a = \operatorname{Tr}_b(\rho) \) invariant. EMIN captures the maximum global ergotropic change induced by locally non-disturbing operations.
	A bipartite state \(\rho\) of subsystems \(A\) and \(B\) is acted upon by the local measurements \( \{\Pi_k^a\} \), yielding the post-measurement state
	\begin{equation}
		\Pi^a(\rho) = \sum_k \Pi_k^a \rho \Pi_k^a.
		\label{ee}
	\end{equation}
	Here, the projectors \( \Pi_k^a \) are taken in the eigenbasis of a non-degenerate local state \( \rho_a =\operatorname{Tr}_b(\rho)\). The difference in ergotropy before and after such a measurement, under a fixed Hamiltonian \(H\), defines EMIN. Following Ref.~\cite{luo2011measurement}, the optimization in Eq.~\ref{e2} can be bypassed by restricting to states where \( \rho_a \) is non-degenerate. We adopt this constraint throughout and analyze the properties of EMIN without the optimization.

	EMIN is related to the \textit{ergotropic gap} \cite{mukherjee2016presence}, defined as the difference between globally and locally extractable work :
	\begin{equation}
		\delta_\rho^\xi = \xi(\rho, H) - \xi^l(\rho, H).
		\label{eg}
	\end{equation}
	Here $\xi^l(\rho, H)=\operatorname{Tr}(\rho H)-\min_{V \in V_l}\operatorname{Tr}(V\rho V^\dagger H)$ with $V_l$ being the set of local unitary operations. Since the reduced states \( \rho_a \) and \( \rho_b \) are invariant under the measurement \( \Pi^a \), the corresponding local ergotropies remain unchanged when the Hamiltonian has no interactions \cite{alimuddin2019bound}. Hence, EMIN can be expressed as
	\begin{equation}
		N_\xi(\rho, H) = \delta_\rho^\xi - \delta_{\Pi^a(\rho)}^\xi,
		\label{emeg}
	\end{equation}
	i.e., the change in ergotropic gap due to the measurement.

	Importantly, the ergotropic gap is sensitive to the form of the Hamiltonian. For instance, consider a system governed by \( H_1 = H_a \otimes \mathbb{I}_b + \mathbb{I}_a \otimes H_b \) versus one with interactions included, \( H_2 = H_1 + H_{\text{int}} \). The ergotropic gap for the same state can differ significantly between \( H_1 \) and \( H_2 \), underscoring the need to account for interactions. While ergotropic gap serves as an indicator of bipartite and multipartite entanglement \cite{puliyil2022thermodynamic,yang2024multiparticle} under \textit{non-interacting} global Hamiltonian, it can not be a valid measure when interactions are present because product states, which are free states of entanglement, can have non-vanishing ergotropic gap under interacting Hamiltonian. To illustrate consider an Hamiltonian whose ground state is an entangled state. If we take our state $\rho$ to be a product state then it can be easily shown that $\xi(\rho,H) > \xi^l(\rho,H)$ since it is impossible to reach the ground state of the Hamiltonian which is entangled by starting with a product state under only local operations i.e., ergotropic gap can be non-vanishing for product states under interacting Hamiltonian. Although the above result shows the limitations of ergotropic gap to serve as an entanglement measure, it offers fresh insights from a quantum thermodynamic perspective. Namely we can conclude that it is possible to extract more work from an uncorrelated state using global operations compared to a local extraction scenario. However we carefully note that the above statement does not necessarily imply such an improvement is always possible. Therefore it is prudent to analyze under what scenarios a quantum advantage can be achieved in ergotropy using quantum resources like nonlocality and global operations.

	This raises a central question:  
	\textit{Under what conditions do the quantum correlations in a given state yield the maximal enhancement in extractable work?}	While ergotropic gap can serve as a possible indicator, optimizing for all possible local unitary operations is a very hard problem and obtaining a closed form solution is not guaranteed \cite{castellano2025parallel}. We show that EMIN generalizes the concept of ergotropic gap to broader settings, providing a framework to explore how the interplay between nonlocal correlations and Hamiltonian structure governs optimal work extraction by deriving a closed form analytical expression for arbitrary bipartite systems and examining its properties.
	
	EMIN offers a novel work-optimization strategy for bipartite systems. It may serve as a diagnostic for identifying quantum advantage in thermodynamic protocols such as quantum batteries and thermal machines, with additional implications in quantum information processing. It is particularly relevant in identifying operations that simultaneously preserve both nonlocality and ergotropy. EMIN also aids in selecting optimal control fields for work extraction, ensuring that the nonlocal content of a quantum state is fully utilized. Randomized sampling over states and Hamiltonians may be employed to identify configurations that maximize ergotropy through EMIN.
	
	\section{Properties}
	If the state is a product state, \( \rho = \rho_a \otimes \rho_b \), then \( \rho = \Pi^a(\rho) \) by construction, and thus \( N_\xi(\rho, H) = 0 \). This follows directly from the definition: tensor product states remain unchanged under locally invariant projective measurements, and therefore yield no ergotropic gain. Unlike ergotropic gap, EMIN is zero for product states under both local and interacting Hamiltonian thereby improving on the shortcomings of ergotropic gap. It is important to note however the converse is not true i.e., a zero EMIN does not imply a product state. This is because EMIN depends not only on the state but also on the Hamiltonian, as becomes evident when considering the class of operations that preserve EMIN. Unlike conventional MIN measures \cite{luo2011measurement,shi2022entanglement,hu2015measurement,muthuganesan2017fidelity}, EMIN is not invariant under all local unitary operations.
	
	\begin{theorem}
		\label{T2}
		\textbf{(Restricted Local Unitary Invariance)}  
		Let \( U = \mathbb{I}_a \otimes U_b \) be a local unitary operator that commutes with the global Hamiltonian \( H \). Then,
		\[
		N_\xi(U \rho U^\dagger, H) = N_\xi(\rho, H).
		\]
	\end{theorem}

	The above result can be proved by considering that ergotropy is invariant under unitary operation that commute with $H$ and imposing further the local measurement commutes with the unitary (see Supplementary Material for proofs of all the Theorems in the main text). Thus EMIN is invariant under a restricted class of local unitaries that do not alter the energy structure imposed by the Hamiltonian. In contrast to standard MIN measures, which are invariant under all local unitaries, the Hamiltonian-dependence of EMIN naturally limits its symmetry. While local unitaries do not affect the intrinsic nonlocal correlations present in the state, the contribution of those correlations to ergotropy may vary depending on their alignment with the Hamiltonian as captured by EMIN. This dependence highlights the thermodynamic character of $N_\xi$ and reflects how the structure of the Hamiltonian mediates the operational utility of nonlocal resources.

	\subsection{EMIN for Pure States:}
	\begin{theorem}
		For an arbitrary pure state $\left| \psi \right\rangle$ with Schmidt decomposition $\sum_{i} \sqrt{\lambda_i} \left| \alpha_i \right\rangle \otimes \left| \beta_i \right\rangle$, and a Hamiltonian $H = \sum_{k} \epsilon_k \left| \epsilon_k \right\rangle \left\langle \epsilon_k \right|$ that admits a Schmidt decomposition $\sum_{l} s_l A_l \otimes B_l$, the EMIN is given by
		\begin{align}
			N_\xi(\rho, H) &= \sum_{i,j,l} \sqrt{\lambda_i \lambda_j} \, s_l \left\langle \alpha_j \right| A_l \left| \alpha_i \right\rangle \left\langle \beta_j \right| B_l \left| \beta_i \right\rangle (1 - \delta_{ij})\nonumber  \\ & \quad + \sum_k \epsilon_k^\uparrow (\lambda_k^\downarrow - \delta_{k0}).
			\label{e3}
		\end{align}
		\label{T3}
	\end{theorem}

	Hence, an explicit expression for EMIN in terms of the Schmidt coefficients of the state and Hamiltonian can be derived, offering analytical insight into how quantum correlations contribute to extractable work. This observation leads to the question of whether the contribution to ergotropy from correlations is always positive. It turns out that contributions from correlations can be non-negative when we impose a certain constraint on the Hamiltonian structure.

	\begin{theorem}
		Let \( H \in \mathbb{H}^a \otimes \mathbb{H}^b \) be a non-interacting Hamiltonian of the form \( H = A \otimes \mathbb{I}^b + \mathbb{I}^a \otimes B \), where \( \mathbb{I}^a, A \in \mathbb{H}^a \) and \( \mathbb{I}^b, B \in \mathbb{H}^b \). Then, for all pure states \( \rho \), the EMIN satisfies $N_\xi(\rho, H)= \sum_k \epsilon_k^\uparrow \lambda_k^\downarrow - \epsilon_0 \ge 0$.
		
		\label{T5}
	\end{theorem}

	\noindent
	This result establishes that, for pure states evolving under non-interacting Hamiltonians, local projective measurements cannot enhance ergotropy. Consequently, any observed ergotropic advantage in such scenarios must stem from entanglement features of the state since $\lambda_k$ are the square of Schmidt coefficients of the state \cite{francica2022quantum}, supporting the interpretation of \( N_\xi \) as a quantifier of thermodynamically relevant entanglement for pure states under local Hamiltonian. Consider the maximally entangled state for which the Schmidt coeffecients are given by $1/\sqrt{d}$ where $d$ is the dimension of the reduced system. For such a state under non-interacting Hamiltonian, EMIN can be written as
	\begin{equation}
		N_\xi(\rho,H)=\frac{1}{d} \left(\sum_{k}\epsilon_k-d \epsilon_0\right).
		\label{mes}
	\end{equation}
	Defining the energy level spacing $s_i=\epsilon_{i+1}-\epsilon_{i}$, Eq. \eqref{mes} can be rewritten as 
	\begin{equation}
		N_\xi(\rho,H)=\frac{1}{d} \left(\sum_{i=0}^{d-1} (d-i)s_i+\epsilon_0\right),
		\label{ls}
	\end{equation}
	leading to dependence on level spacing of Hamiltonian. Thus EMIN also quantifies that under highly non-degenerate non-interacting Hamiltonian, entanglement in the pure state can contribute to higher ergotropy.

	\subsection{EMIN for Mixed States:}
	To evaluate EMIN for a general bipartite state, we consider the following Hilbert-Schmidt basis formalism \cite{luo2011measurement}. An orthonormal basis in $\mathbb{H}$ is given by $(X_i: i = 0, 1, 2, \cdots, n^2 - 1)$ such that $\langle X_i | X_j \rangle = \operatorname{Tr}(X_i X_j^\dagger) = \delta_{ij}$, where $\langle \cdot | \cdot \rangle$ denotes the Hilbert-Schmidt inner product. Consider orthonormal bases $(X_i: i = 0, 1, \cdots, n^2 - 1)$ and $(Y_j: j = 0, 1, \cdots, m^2 - 1)$ for $\mathbb{H}^a$ and $\mathbb{H}^b$, respectively, with $X_0 = \mathbb{I}^a / \sqrt{n}$ and $Y_0 = \mathbb{I}^b / \sqrt{m}$. Then, any bipartite state can be expressed as \cite{luo2011measurement,muthuganesan2017fidelity}:
	
	\begin{align}
		\rho &= \frac{1}{\sqrt{nm}} \frac{\mathbb{I}^a}{\sqrt{n}} \otimes \frac{\mathbb{I}^b}{\sqrt{m}} + \sum_{i=1}^{n^2 - 1} x_i X_i \otimes \frac{\mathbb{I}^b}{\sqrt{m}}\nonumber \\ &\quad  + \frac{\mathbb{I}^a}{\sqrt{n}} \otimes \sum_{j=1}^{m^2 - 1} y_j Y_j + \sum_{i=1}^{n^2 - 1} \sum_{j=1}^{m^2 - 1} t_{ij} X_i \otimes Y_j.
		\label{bp}
	\end{align}
	
	\begin{theorem}
		For any state $\rho$ of the form in Eq.~\ref{bp} and Hamiltonian \( H = \sum_k \epsilon_k |\epsilon_k\rangle\langle \epsilon_k| \), admitting a Schmidt decomposition \( H = \sum_l s_l A_l \otimes B_l \), EMIN is given by
		\begin{widetext}
			\begin{equation}
				N_\xi(\rho, H) = \sum_l \sum_{i=1}^{n^2 - 1} \sum_{j=1}^{m^2 - 1} 
				t_{ij} s_l \left[ \operatorname{Tr}(X_i A_l) - \operatorname{Tr} \left( \sum_k P_k X_i P_k A_l \right) \right] 
				\operatorname{Tr}(Y_j B_l) 
				+ \sum_k \lambda_{\Pi^a_k}^{\downarrow} \epsilon_k^{\uparrow} 
				- \sum_k \lambda_k^{\downarrow} \epsilon_k^{\uparrow},
				\label{bpemin}
			\end{equation}
		\end{widetext}
		
		where \( \{ \lambda_k \} \) and \( \{ \lambda_{\Pi^a_k} \} \) are the eigenvalues of the state before and after local projective measurement, and \( P_k \) are projectors onto the eigenbasis of \( \rho_a \).
		\label{T6}
	\end{theorem}

	This result provides an explicit expression for EMIN in terms of the structural components of the state and Hamiltonian, illuminating how correlations and measurement-induced changes jointly influence extractable work in general mixed states.
	
	\begin{theorem}
		For non-interacting Hamiltonians of the form \( H = A \otimes \mathbb{I}^b + \mathbb{I}^a \otimes B \), it holds that
		$
		N_\xi(\rho, H) = E_p(\Pi^a(\rho)) - E_p(\rho) \ge 0.
		$
		\label{T7}
	\end{theorem}
	
	The above result can be proved using the Schur concavity of passive state energies \cite{marshall2011inequalities}. Since $N_\xi(\rho,H)$ is non negative, it remains a valid measure of correlations for mixed states under non-interacting Hamiltonian. For the case of mixed states which has correlations beyond entanglement $N_\xi(\rho,H)$ is related to the geometric nonlocality which includes correlations due to both entanglement and mixing. 
	Conversely, if local measurements do increase ergotropy in the presence of interactions, it indicates that certain nonlocal correlations, though quantum in origin, can hinder extractable work, highlighting a context-dependent thermodynamic cost of nonlocality. It is also important to note that EMIN quantify correlations under intereacting Hamiltonian, but can indicate the ergotropic usefulness of correlations. This is because of the energy of the state not being the same after measurement due to the Hamiltonian interactions. For instance consider the scenario of the state after measurement having more energy under a certain Hamiltonian. Then subtracting the energies before and after measurement will give a negative contribution to $N_\xi$. Thus it is possible to have $N_\xi <0$ even for entangled states.

	\begin{figure*}[ht]
		\centering
		\includegraphics[width=1\linewidth]{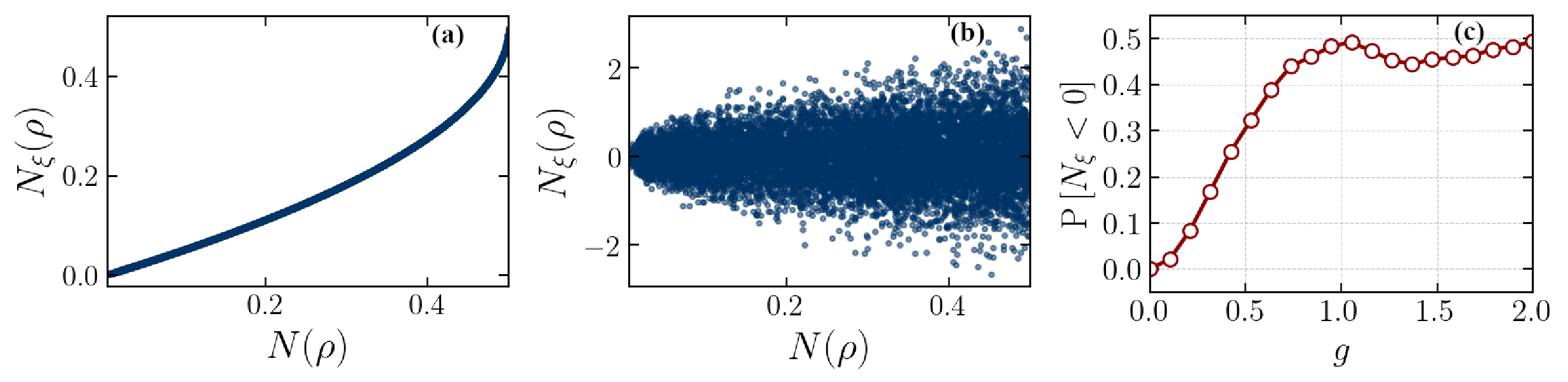}
		\caption{EMIN for hybrid systems under (a) weak ($g=0.05$) and (b) strong interactions ($g=2$) is plotted against the HS norm $N(\rho)$. (c) The probability of negative EMIN as a function of interaction strength $g$. $10^4$ random states are generated to illustrate the results. }
		\label{f3}
	\end{figure*}
	
	For demonstration, consider the Jaynes-Cummings model with single qubit mechanical mode given by the Hamiltonian
	\begin{equation}
		H=a^\dagger a+\frac{\sigma_z}{2}+g(\sigma_+ a+\sigma_- a^\dagger).
	\end{equation}
	Here $g$ is the interaction between mechanical mode given by spin operators $\sigma_i$ and field mode given by $a^\dagger a$. $\sigma_{\pm}$ denotes the spin-ladder operators $\sigma_x \pm i \sigma_y$. To calculate $N_\xi$, $10^4$ random states in which the mechanical mode has the reduced state of the form $\left|\alpha\right|^2 \left|0\right>\left<0\right| +\left|\beta\right|^2 \left|1\right>\left<1\right|$ are generated. The Hilbert-Schmidt norm (HS norm) $N(\rho)=||\rho-\Pi^a(\rho)||^2$ \cite{luo2011measurement}, is used to quantify the geometric measurement induced nonlocality where $||X||^2$ denotes $Tr(X^\dagger X)$.

	For very weak interactions at $g=0.05$ illustrated in Fig. \ref{f3}(a), EMIN remains non-negative and it can be observed that EMIN has a one to one correspondence with the geometric distance as a consequence of Theorem \ref{T7}. The positivity condition for \( E_p(\Pi^a(\rho)) - E_p(\rho) \ge 0 \) holds irrespective of the Hamiltonian structure. Therefore, under interacting Hamiltonians, EMIN can be influenced by changes in the energy expectation value of the state due to measurement. If \( E(\rho) > E(\Pi^a(\rho)) \), the contribution to ergotropy from nonlocal correlations can surpass that of the non-interacting case. Conversely, when
	\[
	E(\rho) + E_p(\Pi^a(\rho)) - E_p(\rho) < E(\Pi^a(\rho)),
	\]
	nonlocal features of the state may reduce extractable work—revealing that certain correlations can act as a thermodynamic hindrance depending on the system's interaction structure. Figure \ref{f3}(b) illustrates the above point where EMIN exhibits a range of positive and negative values for higher interaction strength of $g=2$. Although negative values are possible in highly interactive Hamiltonian, it is crucial to note that $N_\xi$ can take higher values by having a wider spread at maximum $N(\rho)$ compared to weakly or non-interacting case. This dichotomy in strongly interacting Hamiltonian highlight the importance of choice of state and Hamiltonian for optimal work extraction and utilization of quantum nonlocal effects. The probability of negative value of EMIN, $P[N_\xi<0]$ can be calculated as the ratio of number of instances yielding negative EMIN to the total number of instances. From Theorem \ref{T7}, it can be deduced that perturbative values of interaction strength will have low probability of showing negativity. This can be further evidenced by Fig. \ref{f3}(c) where the probability increases with $g$ until it reaches $0.5$ where saturation begins.

	With the above properties of EMIN, two key observations can be made to further elucidate the role of nonlocality and the significance of the Hamiltonian's structure.

	\begin{observation}
		A nonzero value of \( N_\xi(\rho) \) implies the presence of nonlocal correlations in the state \( \rho \). However, the converse does not necessarily hold—there exist states with nonlocal features for which \( N_\xi(\rho) = 0 \). Such states exhibit nonlocality without contributing to nonlocal energy locking. This phenomenon highlights that certain nonlocal states may not offer any energetic advantage via nonlocal mechanisms~\cite{alimuddin2025ergodiscord}.
		\label{ob1}
	\end{observation}
	
	A simple illustration of the above statement can be found in the Supplementary material.

	\begin{observation}
		A negative value of \( N_\xi(\rho) \) indicates that the nonlocal correlations in the state are detrimental to the available ergotropy, effectively suppressing its extractable work potential. This behavior shows a contrast of nonlocal energy locking, where the nonlocal correlations inhibit the state's ability to gain additional ergotropy that would be accessible in their absence.
	\end{observation}

	In general, EMIN can assume positive, negative, or zero values for states with nonlocal correlations. A positive value underscores the constructive role of nonlocal resources in enhancing ergotropy, while a negative value reflects a post-measurement energy gain, signifying that the nonlocal correlations reduce ergotropic capacity under the given Hamiltonian.
	
	To further study the bounds of EMIN, consider the following identity for the Gibbs state $\rho_\beta=e^{-\beta H}/Z$ given by \cite{francica2020quantum}
	\begin{equation}
		D(\rho||\rho_\beta)=\beta \operatorname{Tr}[H(\rho-\rho_\beta)]-S(\rho)+S(\rho_\beta),
		\label{ideq}
	\end{equation}
	where $D(a||b)=Tr[a (log(a)-log(b))]$ is the relative entropy and $S(a)=-\operatorname{Tr}[alog(a)]$ is the Von Nuemann entropy. By using the above identity and imposing the positivity of the relative entropy, analytical bounds for EMIN can be derived as
	
	\begin{align}
		&D(\Pi^a(\rho)^p||\rho_\beta)-D(\Pi^a(\rho)||\rho_\beta)-D(\rho^p||\rho_\beta) \le \beta N_\xi(\rho,H) \nonumber \\  & \ge D(\rho||\rho_\beta)+D(\Pi^a(\rho)||\rho_\beta)-D(\rho^p||\rho_\beta) 
		\label{bdeq}
	\end{align}
	for some finite value of $\beta$. Substituting the constraint of non-interacting Hamiltonian, the bounds in \eqref{bdeq} reduces to the bounds of the discord contribution to ergtotropy which was derived with the condition of finding the difference in ergotropy with the closest classical state which has the same energy has the state itself \cite{francica2022quantum}. 
	
	\section{Summary and Conclusion}
	Viewing measurement-induced nonlocality (MIN) through the lens of ergotropy offers a novel framework for classifying the energetic utility of nonlocal quantum correlations in both interacting and non-interacting Hamiltonian scenarios. Motivated by the pursuit of efficient energy manipulation, quantum thermodynamics has emerged as a powerful paradigm for identifying and classifying quantum resources.
	In this context, EMIN provides an operational framework that captures the “usefulness of nonlocal resources” for work extraction, enabling the optimization of both quantum states and system Hamiltonians. Unlike conventional MIN measures, EMIN is inherently Hamiltonian-dependent and consequently lacks full local unitary invariance, highlighting its thermodynamic relevance. Moreover, EMIN can be computed analytically for arbitrary bipartite states and Hamiltonians, making it a convenient and powerful tool for probing the ergotropic utility of nonlocal resources.
	
	The positivity of EMIN for all non-interacting Hamiltonian demonstrates that nonlocal correlations will inevitably enhance ergotropy in such cases. However for interacting Hamiltonians, the energy content of the state may change post-measurement, and we show that interactions can either amplify or diminish the nonlocal contribution to ergotropy, introducing a rich dichotomy in behavior.
	Our Letter thus advances the understanding of bipartite nonlocality in quantum thermodynamics and lays the groundwork for future investigations into multipartite scenarios and resource-driven thermodynamic protocols.

	\bibliography{cite}
	\onecolumngrid
	\section*{Supplementary material}
	\subsection{Proof of theorem 1}
	\begin{proof}
		We begin with the following corollary:
		
		\begin{corollary}
			\label{cor1}
			Ergotropy \( \xi(\rho, H) \) is invariant under unitary operations \( U \) such that \( [U, H] = 0 \).
		\end{corollary}
		
		This follows directly from the definition of ergotropy:
		\begin{align}
			\xi(U \rho U^\dagger, H) 
			&= \operatorname{Tr}(U \rho U^\dagger H) - \min_V \operatorname{Tr}(V U \rho U^\dagger V^\dagger H) \nonumber \\
			&= \operatorname{Tr}(\rho H) - \min_W \operatorname{Tr}(W \rho W^\dagger H) = \xi(\rho, H),
			\label{t2e1}
		\end{align}
		where \( W = VU \) is unitary, and the commutation condition \( [U, H] = 0 \) ensures the invariance of the energy terms.
		
		Now consider \( N_\xi(U \rho U^\dagger, H) = \xi(U \rho U^\dagger, H) - \xi(\Pi^a(U \rho U^\dagger), H) \). To show equality with \( N_\xi(\rho, H) \), we require:
		\[
		\xi(U \rho U^\dagger, H) = \xi(\rho, H)  \] and \[ \xi(\Pi^a(U \rho U^\dagger), H) = \xi(\Pi^a(\rho), H).
		\]
		
		The first condition follows from Corollary~\ref{cor1}. For the second, note that for \( U = \mathbb{I}_a \otimes U_b \), the measurement projectors \( \{\Pi_k^a = |k\rangle\!\langle k| \otimes \mathbb{I}_b\} \) commute with \( U \), and thus
		\begin{equation}
			\Pi^a(U \rho U^\dagger) = U \Pi^a(\rho) U^\dagger.
			\label{t2e2}
		\end{equation}
		Because \( U \) also commutes with \( H \), the ergotropy of \( \Pi^a(\rho) \) is preserved under conjugation by \( U \). Hence,
		\[
		\xi(\Pi^a(U \rho U^\dagger), H) = \xi(\Pi^a(\rho), H),
		\]
		which completes the proof.
	\end{proof}
	\subsection{Proof of theorem 2}
	\begin{proof}
		For a pure state,
		\begin{equation}
			\rho = \sum_{i,j} \sqrt{\lambda_i \lambda_j} \left| \alpha_i \right\rangle \left\langle \alpha_j \right| \otimes \left| \beta_i \right\rangle \left\langle \beta_j \right|,
			\label{t3e1}
		\end{equation}
		the state after projective measurement on subsystem $a$ is
		\begin{equation}
			\Pi^a(\rho) = \sum_{k} \lambda_k \left| \alpha_k \beta_k \right\rangle \left\langle \alpha_k \beta_k \right|.
			\label{t3e2}
		\end{equation}
		The ergotropy before measurement takes the form
		\begin{equation}
			\xi(\rho, H) = \sum_{i,j,l} \sqrt{\lambda_i \lambda_j} \, s_l \left\langle \alpha_j \right| A_l \left| \alpha_i \right\rangle \left\langle \beta_j \right| B_l \left| \beta_i \right\rangle - \epsilon_0.
			\label{t3e3}
		\end{equation}
		The ergotropy after measurement is
		\begin{equation}
			\xi(\Pi^a(\rho), H) = \sum_{k,l} \lambda_k \, s_l \left\langle \alpha_k \right| A_l \left| \alpha_k \right\rangle \left\langle \beta_k \right| B_l \left| \beta_k \right\rangle - \sum_k \epsilon_k^\uparrow \lambda_k^\downarrow,
			\label{t3e4}
		\end{equation}
		where the eigenvalues $\epsilon_k$ and $\lambda_k$ are arranged in increasing and decreasing order, respectively, to ensure minimization. Subtracting Eq.~\ref{t3e4} from Eq.~\ref{t3e3} establishes Theorem~\ref{T3}.
	\end{proof}
	\subsection{Proof of theorem 3}
	\begin{proof}
		Using the linearity of the trace and writing the pure state \( \rho = \ket{\psi}\bra{\psi} \) in its Schmidt basis as \( \ket{\psi} = \sum_k \sqrt{\lambda_k} \ket{\alpha_k} \otimes \ket{\beta_k} \), the ergotropy under \( H = A \otimes \mathbb{I}^b + \mathbb{I}^a \otimes B \) becomes:
		\begin{align}
			\xi(\rho, H) &= \sum_{i,j} \sqrt{\lambda_i \lambda_j} 
			\left( \bra{\alpha_j} A \ket{\alpha_i} \bra{\beta_j} \mathbb{I} \ket{\beta_i} \right. \nonumber \\
			&\quad \left. + \bra{\alpha_j} \mathbb{I} \ket{\alpha_i} \bra{\beta_j} B \ket{\beta_i} \right) - \epsilon_0 ,
			\label{t5e1}
		\end{align}
		
		and the ergotropy after the measurement is
		\begin{align}
			\xi(\Pi^a(\rho), H) &= \sum_k \lambda_k \left( \bra{\alpha_k} A \ket{\alpha_k} \bra{\beta_k} \mathbb{I} \ket{\beta_k} +\right. \nonumber \\& \quad \left. \bra{\alpha_k} \mathbb{I} \ket{\alpha_k} \bra{\beta_k} B \ket{\beta_k} \right) - \sum_k \epsilon_k^\uparrow \lambda_k^\downarrow. \label{t5e2}
		\end{align}
		
		Since \( \Pi^a(\rho) \) retains the same marginal spectra as \( \rho \), and the Hamiltonian is non-interacting, the average energy before and after the measurement remains unchanged. Thus, the difference in ergotropy arises solely due to the change in the passive state contribution:
		\begin{align}
			N_\xi(\rho, H) = \sum_k \epsilon_k^\uparrow \lambda_k^\downarrow - \epsilon_0. \label{t5e5}
		\end{align}
		
		By definition, \( \epsilon_0 \) is the minimal energy attainable by any state under the Hamiltonian \( H \). Therefore, the difference is always non-negative, completing the proof.
	\end{proof}
	
	\subsection{Proof of theorem 4}
	\begin{proof}
		The expression for EMIN can be rewritten as:
		\begin{equation}
			N_\xi(\rho, H) = \operatorname{Tr}\left[(\rho - \Pi^a(\rho)) H\right] + E_p(\Pi^a(\rho)) - E_p(\rho).
			\label{t6e1}
		\end{equation}
		
		From Eq.~\ref{bp}, the marginal state is given by:
		\begin{equation}
			\rho_a = \frac{\mathbb{I}^a}{n} + \sqrt{m} \sum_{i=1}^{n^2 - 1} x_i X_i,
			\label{t6e2}
		\end{equation}
		since \( \operatorname{Tr}(Y_j) = 0 \) for all \( j \ne 0 \). For a locally invariant projective measurement, we have:
		\[
		\sum_{i=1}^{n^2 - 1} x_i X_i = \sum_k P_k \left( \sum_{i=1}^{n^2 - 1} x_i X_i \right) P_k.
		\]
		Then, the difference between the pre- and post-measurement states is:
		\begin{equation}
			\rho - \Pi^a(\rho) = \sum_{i=1}^{n^2 - 1} \sum_{j=1}^{m^2 - 1} t_{ij} \left( X_i - \sum_k P_k X_i P_k \right) \otimes Y_j.
			\label{t6e3}
		\end{equation}
		Substituting Eq.~\ref{t6e3} into Eq.~\ref{t6e1} and applying the decomposition of \( H \) leads to Eq.~\ref{bpemin}, after straightforward algebraic manipulation.
	\end{proof}
	\subsection{Proof of theorem 5}
	\begin{proof}
		Substituting the non-interacting Hamiltonian into Eq.~\ref{bpemin}, we obtain:
		
		\begin{equation}
			\begin{aligned}
				N_\xi(\rho, H) = & \sum_{i=1}^{n^2 - 1} \sum_{j=1}^{m^2 - 1} t_{ij} \Big( \left[ \operatorname{Tr}(X_i A) - \operatorname{Tr}\left( \sum_k P_k X_i P_k A \right) \right] \operatorname{Tr}(Y_j) + \left[ \operatorname{Tr}(X_i) - \operatorname{Tr}\left( \sum_k P_k X_i P_k \right) \right] \operatorname{Tr}(Y_j B) \Big) \\
				&+ E_p(\Pi^a(\rho)) - E_p(\rho).
			\end{aligned}
			\label{t7e1}
		\end{equation}

		Since \( \operatorname{Tr}(X_i) = \operatorname{Tr}(Y_j) = 0 \) for all \( i,j \ne 0 \), all trace terms in the double summation vanish, leaving:
		\begin{equation}
			N_\xi(\rho, H) = E_p(\Pi^a(\rho)) - E_p(\rho).
			\label{t7e2}
		\end{equation}
		
		Let the passive energy \( E_p \) be represented by the symmetric function \( F(\{\lambda_k\}) = \sum_k \lambda_k^\downarrow \epsilon_k^\uparrow \). This function satisfies Schur concavity\cite{marshall2011inequalities}, i.e.,
		\begin{equation}
			(\lambda_i - \lambda_j)\left( \frac{\partial F}{\partial \lambda_i} - \frac{\partial F}{\partial \lambda_j} \right) \le 0.
			\label{t7e3}
		\end{equation}
		The post-measurement state \( \Pi^a(\rho) = \sum_k (\Pi_k^a \otimes \mathbb{I}^b) \rho (\Pi_k^a \otimes \mathbb{I}^b) \) defines a unital map, the quantum analogue of a doubly stochastic map \cite{preskill2015lecture}, and thus \( \Pi^a(\rho) \prec \rho \) where the symbol $\prec$ denotes majorization \cite{marshall2011inequalities}. By the Schur concavity of \( F \), we have \( E_p(\Pi^a(\rho)) \ge E_p(\rho) \), completing the proof.
	\end{proof}
	
	\subsection{Illustration of Observation-1}
	Consider the pure state \( \left|\psi \right\rangle = \alpha \left|00\right\rangle + \beta \left|11\right\rangle \), where \( \alpha \in \mathbb{R} \setminus \{0\} \) and \( \beta = \sqrt{1 - |\alpha|^2} \neq 0 \). The pre- and post-measurement states are given by
	\[
	\rho = \alpha^2 \left|00\right\rangle\left\langle 00\right| + \beta^2 \left|11\right\rangle\left\langle 11\right| + \alpha\beta \left( \left|00\right\rangle\left\langle 11\right| + \left|11\right\rangle\left\langle 00\right| \right),
	\]
	and
	\[
	\Pi^a(\rho) = \alpha^2 \left|00\right\rangle\left\langle 00\right| + \beta^2 \left|11\right\rangle\left\langle 11\right|.
	\]
	The Hilbert-Schmidt norm between these two states is \( \| \rho - \Pi^a(\rho) \|^2 = 2 \alpha^2 \beta^2 \), indicating the presence of nonlocal correlation for nonzero \( \alpha, \beta \).
	
	Now consider the Hamiltonian \( H = \sigma_x \otimes \sigma_z \), which has eigenvalues \( -1, -1, 1, 1 \). Evaluating the ergotropy, we find
	\[
	\xi(\rho) = \mathrm{Tr}(\rho H) - \mathrm{Tr}(\rho^p H) = 1,
	\quad \] and \[ \quad
	\xi(\Pi^a(\rho)) = \mathrm{Tr}(\Pi^a(\rho) H) - \mathrm{Tr}(\Pi^a(\rho)^p H) = 1.
	\]
	Thus, \( N_\xi = \xi(\rho) - \xi(\Pi^a(\rho)) = 0 \), despite the state's evident nonlocality. This example illustrates the essence of Observation~\ref{ob1}: the nonlocal correlations present in \(  \left|\psi \right\rangle \) are inactive in terms of contributing to ergotropy under the chosen Hamiltonian \( H \).
\end{document}